# Big Models: From Beijing to the whole China

1 Ying Long, Beijing Institute of City Planning
2 Kang Wu, Capital University of Economics and Business, China
3 Jianghao Wang, Chinese Academy of Sciences, China
4 Zhenjiang Shen, Kanazawa University, Japan

**Abstract:** This paper propose the concept of big model as a novel research paradigm for regional and urban studies. Big models are fine-scale regional/urban simulation models for a large geographical area, and they overcome the trade-off between simulated scale and spatial unit by tackling both of them at the same time enabled by emerging big/open data, increasing computation power and matured regional/urban modeling methods. The concept, characteristics, and potential applications of big models have been elaborated. We addresse several case studies to illustrate the progress of research and utilization on big models, including mapping urban areas for all Chinese cities, performing parcel-level urban simulation, and several ongoing research projects. Most of these applications can be adopted across the country, and all of them are focusing on a fine-scale level, such as a parcel, a block, or a township (sub-district), which is not the same with the existing studies using conventional models that are only suitable for a certain single or two cities or regions, or for a larger area but have to significantly sacrifice the data resolution. It is expected that big models will mark a promising new era for the urban and regional study in the age of big data.

## 1 A golden era of Big Models

Applied regional/urban models have attracted extensive attention from researchers in recent decades. Regional models are used for regional analysis at a macro-geographic level, such as for a collection of cities or an entire country. They generally involve a variety of spatial analysis approaches and statistical methods. Broadly speaking, regional models are data processing and analysis-oriented rather modeling per se. Generally, there are a basket of modeling and simulation approaches in urban models (Batty, 2009). They are commonly used for understanding and predicting urban systems through abstracting and generalizing different components of a city. Urban models were first developed in the early 1950s and moved through several phases as they developed and evolved. Figure 1 presents the development line of urban models from static to dynamic models. The dynamic models further include top-down differential equation-based models using system dynamics and currently prevailing bottom-up models using cellular automata or agent-based approaches. The spatial unit of urban models is also in a transition from a larger territorial unit such as a large grid or a zone to a smaller unit such as a block, a parcel, or a building (Hunt et al, 2005; Wegener, 2004). Generally, these two types of models are utilized separately. According to existing research on applied regional/urban models, they are rarely used simultaneously or synthetically.

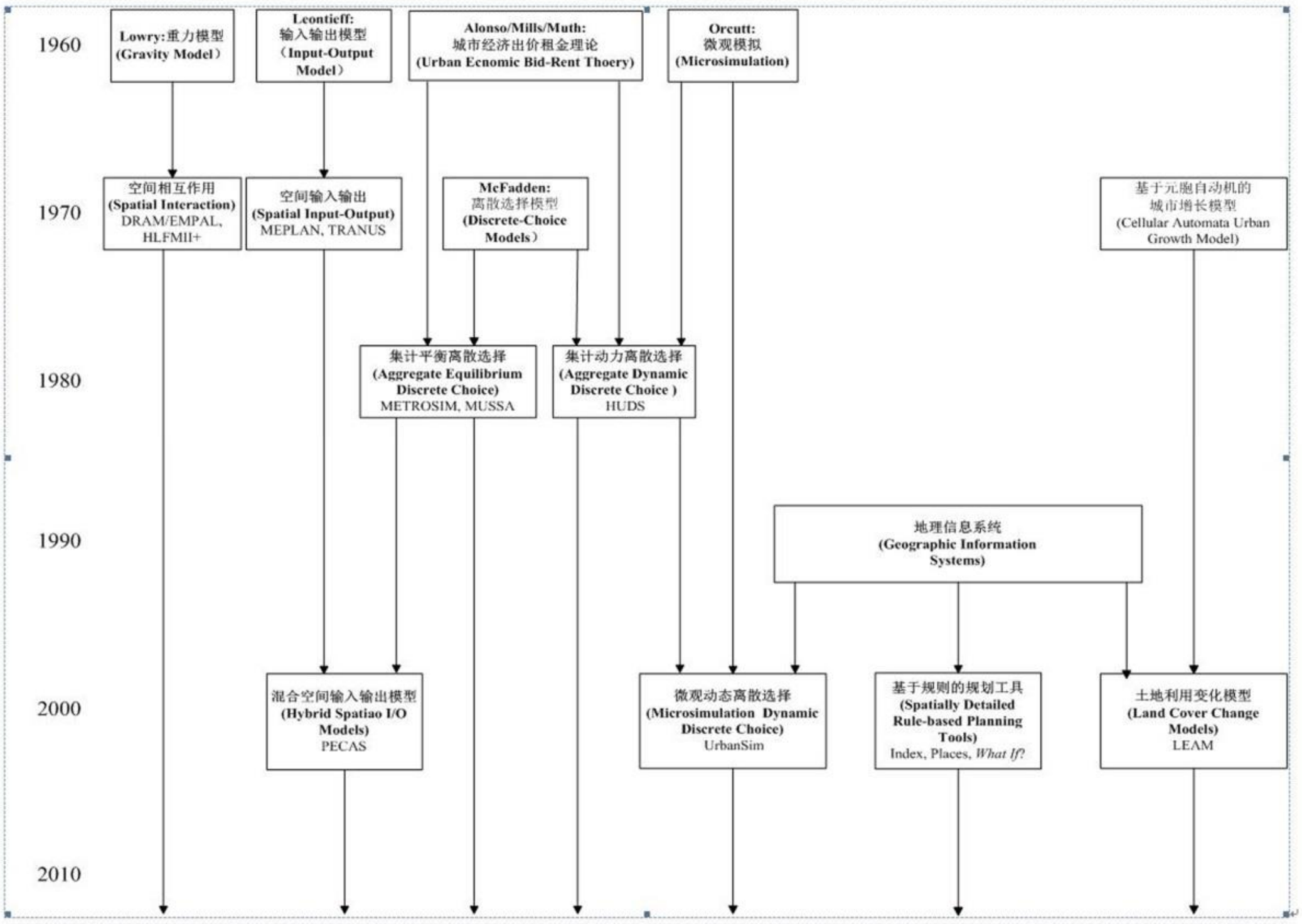

Figure 1 The development line of Applied Urban Models
(Adapted from Paul Waddell, Dynamic Microsimulation: UrbanSim, Webinar 5 of 8-part TMIP, Webinar series on land use forecasting methods)

In practice, the existing applied regional/urban models can fall into two clusters based on their geographical scale and spatial unit. One is a fine-scaled model for a small area, e.g. part of a city or an entire city. The modeling spatial unit can be a parcel, a block, or a small cell. The other is a model for a large area, such as a region or an entire country. The modeling unit can be a county or a super cell. Because there is a general tradeoff between the spatial extent and the resolution of baseline data for modeling due to the data paucity, it is hard to develop a model that can be applied to a large geographic area but with a small spatial unit (see Figure 2).

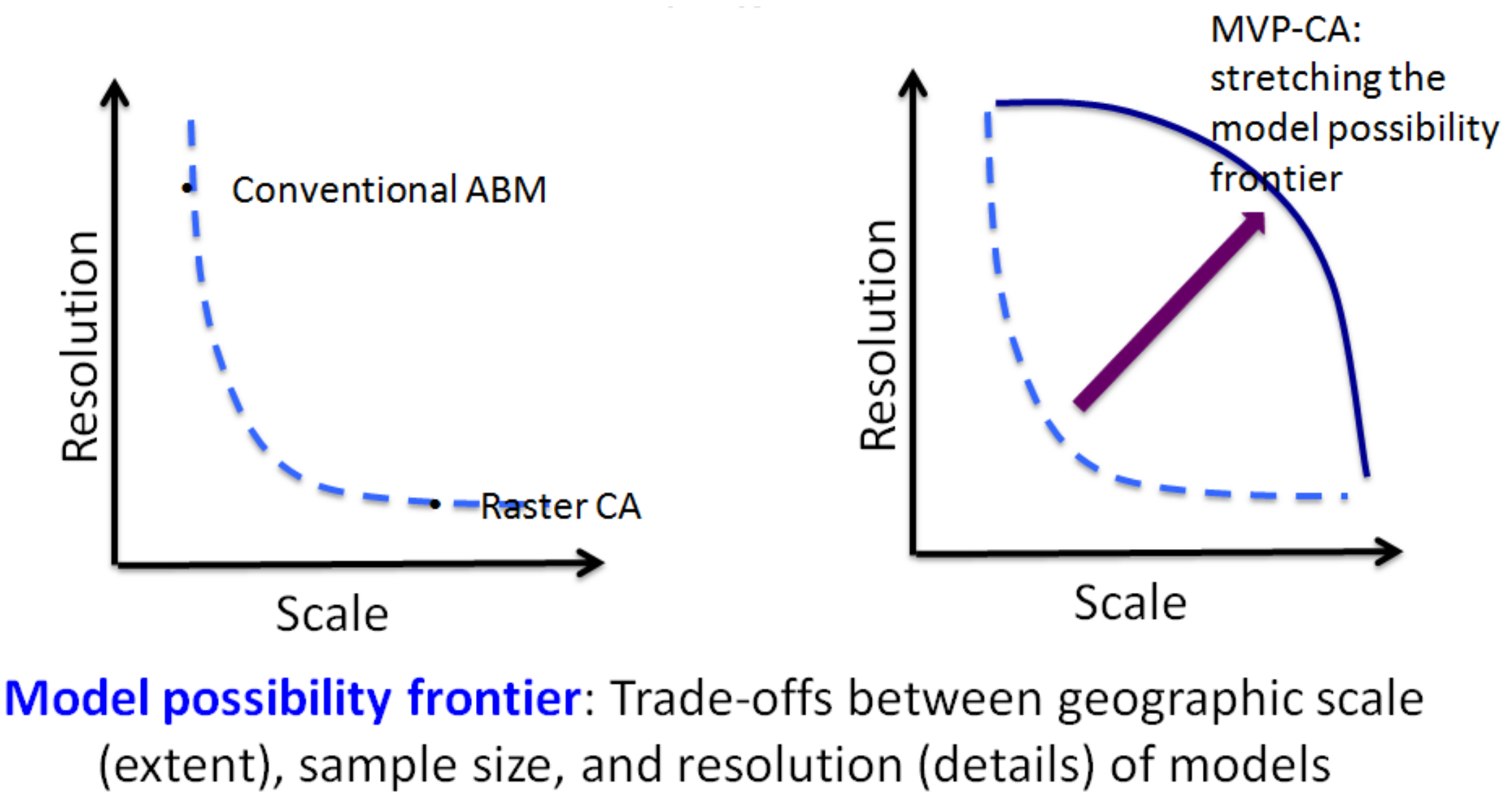

Figure 2 Conventional models vs. "Big Models" (MVP-CA is our first big model for simulating urban expansion at the parcel level for all Chinese cities)

To the best of our knowledge, fine-scale applied urban models for a large area have been rare in academic research. As explained earlier, this stems largely from the lack of data and computation capacity limitation, which are particularly true in the case of China. More often than not the actual datasets required to conduct such analyses did not exist or were hard to collect or obtain. In addition, collecting fine-scale data for feeding models in medium- and small-sized cities is often constrained by poorly developed digital infrastructures. This condition, to some degree, has obstructed the progress of fine-scale urban simulation for a large area in developing countries in general and in China in particular. Overcoming data shortfalls has become the top priority for fine-scale urban simulation in developing countries, even in some developed countries.

In this article, the fine-scale urban simulation model for a large geographical area is termed a "big model". Big models are data-driven urban simulation tools involving a variety of modeling approaches. As a new type of research paradigm for urban and regional studies, it overcomes the trade-off between simulated scale and spatial unit by tackling both of them at the same time. More importantly, as our ability to collect, store, and process data has increased remarkably in recent years since the digital revolution, big models would provide us with new opportunities for better understanding how cities work. There are four major reasons making the widespread use of big models happen. (1) Today, big data, such as mobile traces, public transport smartcard records, online check-ins/points-of-interest, and floating car trajectories, are becoming pervasively available. The spread of mobile technologies and personal computing has made generating, tracking, and recording individual data part of daily life, greatly supporting the analysis and modeling with rich datasets. Some scholars even advocate that data are models themselves (Batty, 2012). (2) Open access to data has been improved significantly as there have been calls for governmental transparency and accountability. For instance, people can access the dataset inventory of planning permits from the official website of Beijing Planning Commission, land transaction records from Beijing Land Bureau, and housing projects from Beijing Housing and Construction Commission. Generally, these records are associated with detailed project-level information, including fine-scale physical characteristics and urban development status. Supported by online geocoding services, these records can be utilized in big models in the form of point datasets. Without painstaking efforts towards an "open government", no such things would have been possible in China. (3) Computational capacity has been largely improved for running big models by means of techniques like parallel computation and Hadoop. (4) For those bottom-up simulation methods adopted by big models, such as cellular automata, agent-based modeling, and network analysis, they have evolved and matured, allowing more sophisticate and powerful application of big models. Therefore, we argue that big models will mark a promising new era for the urban and regional study field.

The purpose of this chapter is to summarize the progress our existing research makes on the application of big models in China. The next section elaborates the basic ideas and characteristics of big models. Section 3 reviews the methodology development and several case studies in utilizing big models on various urban and regional researches. In the end, we conclude with a summary of our findings and suggest directions for further research.

## 2 Big Models: A novel research diagram for urban and regional studies

Big models have the following characteristics. First, they need large-scale geographic data but are not limited to the so called "big data" for initialization. The data may be collected at the individual observation level or based on small geographical units. Second, both the existing inter-urban and intra-city analysis methods can be integrated in big models (see Figure 3 for an illustration of a big model combining inter-city and intra-urban approaches). Third, the geographic extent of big models is generally larger than that of conventional models but with similar spatial scale of simulation units. For instance, quality-of-life (QOL) studies can draw conclusions on a city using data at the block/parcel level. But with big models, the analysis of QOL can be conducted to a larger geographic area, such as for a region or an entire country, and still maintain the same spatial resolution. Fourth, for a same geographical area, a big model can achieve a higher spatial resolution when compared to a conventional model. A good example is that, in a national-scale population density research, the conventional models may only be applicable at the county or city level, whereas big models driven by fine-scale datasets make it possible to address the issue at the sub-district or block level, thus helping bring out more meaningful implication for urban spatial planning and policies.

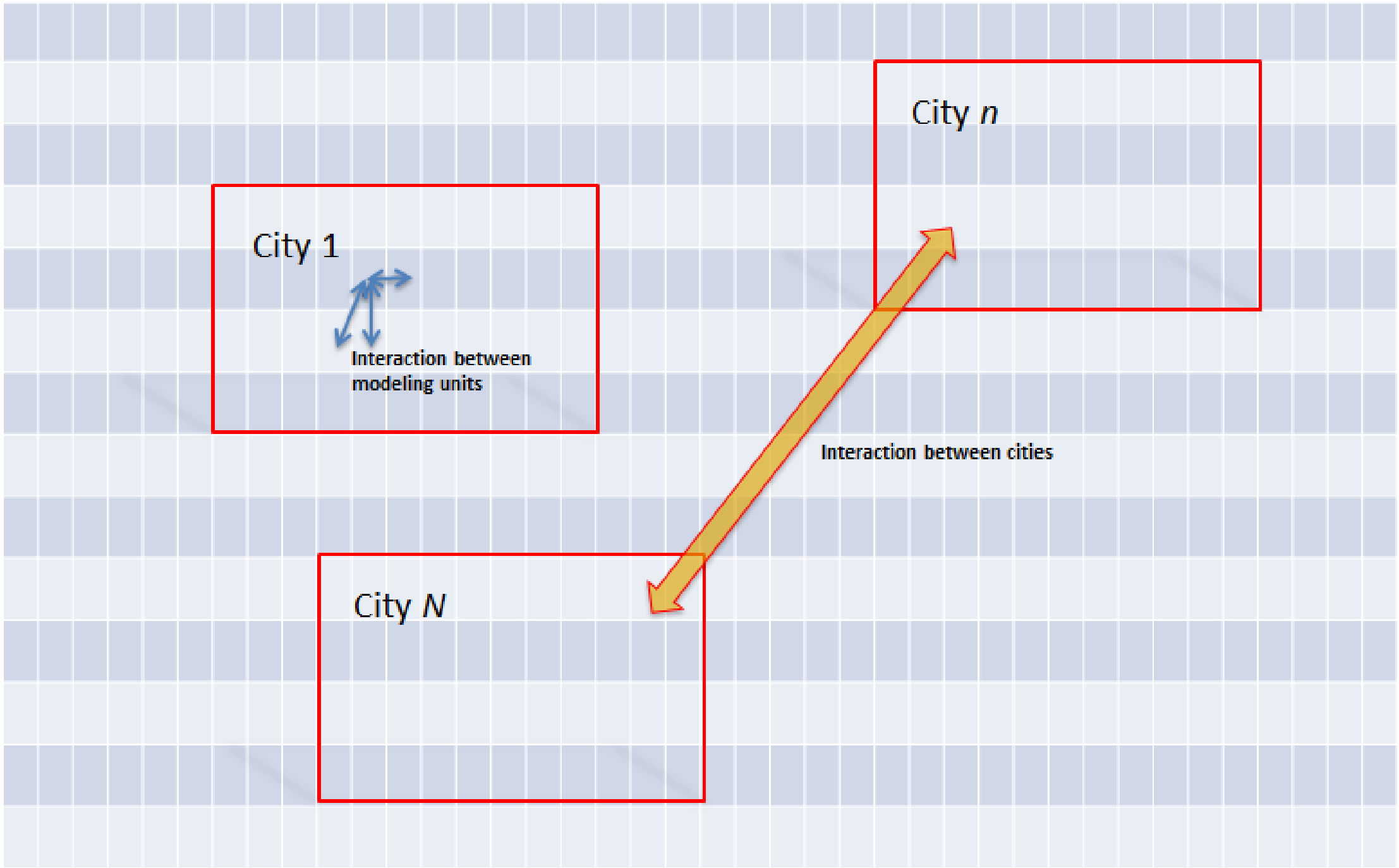

Figure 3 An illustration of a big model integrating intra-and inter-urban methodologies

Big models can be applied in the following avenues. First, all sized cities' urban dynamics can be detected via using big models, in the context that most of applied urban models (AUMs) are for big cities where data infrastructure and technical capacity are much better than middle- and small-sized cities, especially in China. Via introducing big models in the national scale studies, the digital divide in data infrastructure and AUMs can be somewhat mitigated. Second, big models featuring fine-level analysis and modeling can well address issues in urban planning, which is in its turn to "planning for people" from its sights on physical world in China, a place where AUMs are in the macro or meso-scales. Third, urban form and network indicators for each city involved in big models can be calculated as new factors, which, together with extensively adopted socio-economic aggregated indicators, are possible to be combined for intra-urban regional analysis. It was not easy to compute these new indicators before due to being lacking of necessary road network and parcel geometries for a lot of cities at the same time. By means of big models, universal laws on how urban form and network configurations governing urban dynamics and performance are expected to find fed with a large amount of cities as samples.

Simulating regional and urban dynamics using fine-scale and large-area big models is promising as follows. (1) Parcels, as clear behavioral units, would be more appealing to local decision makers and citizens, since each parcel rather a regular cell has a fixed boundary bounded with local images and knowledge; (2) Land use regulations could be distributed to parcels directly, and each city would have access to the simulation results. This would benefit those cities with no financial or intelligent stock supervising or being aware of future developments through our model.; (3) The simulation results could be compared in the city level in the fine scale so that some intra-city phenomenon could be observed. (4) Such model enables integrating spatial interaction analysis (flows and networks) for future. We argue that the research on big models is a scientific question that for solving scientific problems.

## 3 Case studies

Our efforts on the development and application of big models represent a first step towards a better understanding of how cities operate and develop using the emerging big data processing and analysis techniques in China. We outline our methodology development and research process of big models with several completed and ongoing research projects. As most of our case studies draw upon online data sources, the methodologies proposed in this chapter can be easily extended and applied to other cases. (Not relevant)

### 3.1 Mapping urban area for all Chinese cities at the parcel/block level

One typical problem for urban studies is how a city can be defined properly (Zipf, 1949; Krugman 1996; Soo, 2005; Batty, 2006). Urban built-up areas play a strong role in representing urban spatial development for planning decisions, management, and urban studies. They not only illustrate spatial patterns, such as the development levels and scales of the built environment, but also reveal socio-economic characteristics within the built-up areas, e.g., population aggregation, social interaction energy consumption, and land use efficiency, thereby reflecting how a city evolves in a complex manner (Batty, 2011). Conventional methods of capturing the borders of a built-up area from the top down have been applied in major cities around the world on a large scale. However, such methods cannot be applied to most of cities in developing countries due to lack of high-resolution data (Long, 2013). Moreover, the research approach of the existing methods for fine-scale studies is conditioned by the presence of data and study context and hence varies from case to case. Against this backdrop, an automatic bottom-up approach was developed in this chapter. Built upon morphological and functional characteristics determined by street network as well as point of interests (POIs), the proposed approach creates a unified way to define fine-scale cities of all sizes.

Though the definitions and measurements of urban built-up areas have been varied, they generally can be classified into three conceptual types: administrative regions, entities-built areas, and functional districts. Urban built-up areas in the United States are defined as Urbanized Areas (UA) in a typical administrative model for spatial statistics containing the incorporated places and census designated places in central places and urban fringes controlling for the population density (Morris et al., 1999). A counterpart in Japan is called "Densely Inhabited District" (DID). DID is a district which has a population density of more than 4000 people per $km^2$. Furthermore, Urban Areas (UA) in UK are derived from entities-built areas where certain real-estate densities are detected through satellite images (Hu et al., 2008). On the other hand, socio-economic factors are also adopted to describe the actual urban areas, e.g. labor force markets and commuter sheds are utilized to represent Metropolitan Area (MA) (Berry, et, at., 1969). Urban built-up areas are defined for different purposes with respect to population characteristics, economic status, and built environments attributes. In order to define the urban extent more explicitly, the examination and integration of morphological and functional characteristics (i.e. demographics and socio-economics) become essential and should be fully taken into account.

There are many ways of recording and mapping urban built-up areas. From the perspective of capturing morphological characteristics, an increasing attention has been focused on remote sensing images and street network. Remote sensing and night-time satellite imaging help us gauge urban activity and measure the extent and shape of built-up areas through capturing land cover information and interpreting light data (Henderson, 1976; Henderson et. al, 2003, He et al., 2006). Apart from that, a number of indicators of street network have been introduced to describe the spatial layout of the built environment and predict their correlation with social effects. Examples are street intersection density (Masucci et al., 2012; Batty et al., 2012), fractal indices (Shen, 2002; Jiang and Yin, 2013), integration, and accessibility. In terms of the functional characteristics, socio-economic statistics such as demographic densities (Rozenfeld et al., 2008), effective employment density (SGS Economics & Planning, 2011, 2012), and infrastructures accessibilities (Hu et al., 2008) have emerged as a standard method of defining urban statistical areas (Champion et al, 2007; US Census Bureau, 2009).

Nevertheless, these approaches have some drawbacks. 1) The processing and interpretation of remote sensing data could be time consuming and costly. The contexts of study areas are varied and thus it is difficult to set proper selection criteria. Also, resolution of satellite imagery is too coarse for detailed mapping and for distinguishing small areas. 2) Geometrical approach could be powerful for the description and analysis of spatial configurations given places, but would be weak for defining or establishing a spatial unit as parameters are various and less explicit. 3) Fine-scale statistics are time-sensitive. But it takes time and money to collect and process data, thus becoming less reliable in response to subtle population and economy-related altering.

In light of this situation, this chapter employs an automated framework – "automatic identification and characterization of parcels (AICP)" – that was proposed by Long and Liu (2013) to delineate urban built-up areas at the parcel level, based on increasingly standardized roadway asset data from ordnance surveys and crowd-sourced point-of-interests (POIs) data. Roadway data are used to identify and describe parcel configuration, and POIs are processed to infer the intensity, function, and mixing of land use and human activities.

The working definition of a parcel is a continuously built-up area bounded by roads. Identifying land parcels and delineating road space are therefore dual problems. In other words, our approach begins with the delineation of road space, and individual parcels are formed as polygons bounded by roads. The delineation of road space and parcels is performed as follows: (1) All roadway data are merged as line features in a single data layer; (2) individual road segments are trimmed with a threshold of 200m to remove hanging segments; (3) individual road segments are then extended on both ends for 20m to connect adjacent but non-connected lines; (4) road space is generated as buffer zones around road networks. A varying threshold ranging between 2-30 m is adopted for different road types (e.g., surface condition, as well as different levels of roads); (5) parcels are delineated as the space left when road space is removed; and (6) a final step involves overlaying parcel polygons with administrative boundaries to determine whether individual parcels belong to a certain administrative unit. Parameters used in these steps are determined pragmatically with topological errors of roadway data in mind.

We define land use density as the ratio between the counts of POIs in/close to a parcel to the parcel area. We further standardized the density to range from 0 to 1 for better inter-city and intra-city density comparison using the following equation: standardized density = log(raw)/log(max), where raw and max correspond to density of individual parcels and the nation-wide maximum density value[1]. We also note that other measures (e.g., online check-ins and floor area ratio) can substitute POIs and approximate the intensity of human activities.

A vector cellular automata (VCA) model is adopted to identify urban parcels from all generated parcels. In this model, each parcel is assigned a value of 0 (urban) or 1 (non-urban). Initially, all parcels are assumed to be rural. To determine the actual status of each parcel, we should take into account not only the individual parcel's intrinsic attributes, such as population density, neighborhood attributes, and some other spatial variables, but also the status of neighboring parcels. The model stops at the iteration when the total area of urban parcels reaches total urban land.

We applied this approach to map city boundaries for all Chinese cities and compared them with urban areas identified by GLOBCOVER, DMSP/OLS and population density. The simulation process and results highlight our proposed framework is more straightforward, time-saving and precise than conventional methods (see Figure 4 for the results in typical cities).

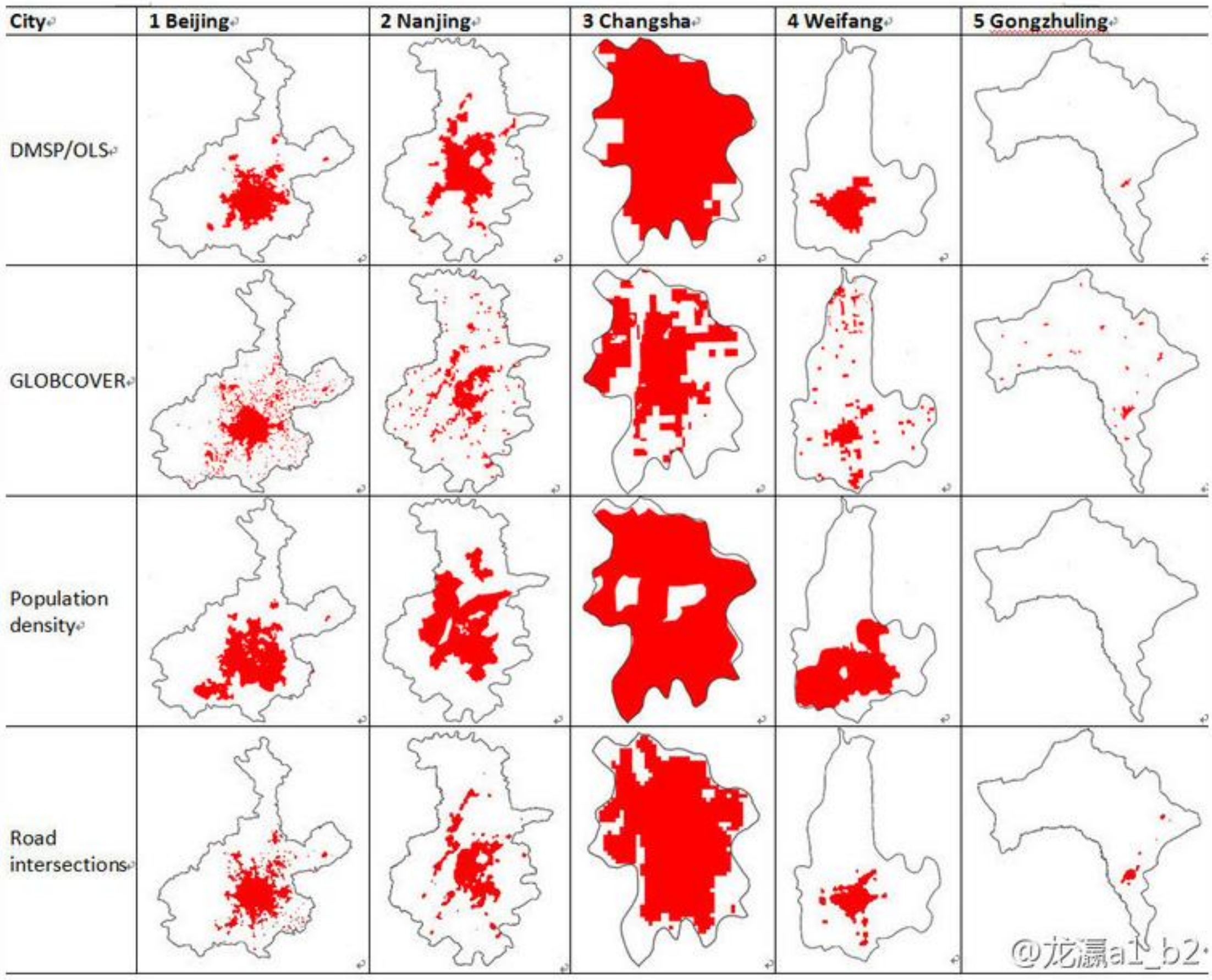

[1] The unit is the POI count per km$^2$. For parcels with no POIs, we assume a minimum density of 1 POI per km$^2$.

Figure 4 Mapped urban areas in five typical Chinese cities by various methods

The contribution of this project lies in three major aspects: data, methodology, and innovation. Firstly, the final product of this project is a database containing urban built-up area maps with detailed parcel features for 654 Chinese cities. Featured by fine-scale parcel information, the detailed road network and POIs datasets consolidated in this research can be applied to support a variety of planning and urban studies projects covering a wide range of geographic extent. Secondly, our research proposed a powerful and consistent approach to identifying urban built-up areas across the country. Unlike previous methods that are somewhat laborious and subjective, our proposed methodology driven by VCA modeling is automatic, straightforward, and objective. The generated parcels can serve as basic spatial units for incorporating other high-resolution ubiquitous and spatially referenced data. In addition to the contribution of delineating urban built-up areas, this research also provides a robust framework for understanding complex urban system across cities from a bottom-up perspective.

## 3.2 Simulating parcel level urban expansion for all Chinese cities

China, as the largest developing country in the world, has experience rapid levels of urbanization in recent year since the introduction of Chinese Reform and Opening-up policies (Montgomery et al, 2008; Liu et al, 2012). Featured by the history's largest flow of rural-to-urban migration and unprecedented economic growth, the urbanization process has shaped and transformed China from a rural to a more urban society. In light of this situation, increasing efforts on urban development assessment and management tools have been made in an attempt to promote a more sustainable development in China; among them are scenario-based urban simulation models (Zhang et al, 2011).

Large-scale simulation models are generally associated with large modeling units in space, like counties or super grids, sometimes reaching tens of square kilometers. Few applied urban models have the ability to pursue a large-scale extent with fine-level units simultaneously due to data paucity and computation capacity limitation as discussed previously. Despite the existing difficulty and infeasibility, urban expansion simulation at a large geographic extent with a fine-scale (i.e. parcel scale) modeling unit could be promising for several reasons. Firstly, simulation and analysis at the parcel level would be more meaningful for local planners, decision makers, and residents to understand, administer, and monitor urban developments. Secondly, simulation modeling at the large geographic extent enables those administrative entities who have limited capacity to analyze and forecast the urban growth taking place within their boundaries by their own to have an insight on overall urban development scenario within the region and to gauge their growth and take action properly. Also, such simulation models make intra-city comparison possible and their results consistent.

In this section, we developed a mega-vector-parcels cellular automata model (MVP-CA) for simulating urban expansion in the parcel level for all 654 Chinese cities. Three modules, the macro module, the parcel generation module, and the vector CA module, were included in the MVP-CA, as shown in Figure 5. The macro module was responsible for setting urban expansion rate in the next five years for each city, taking into account historical urban expansion rate and national spatial development strategies. The parcel generation module was used for identifying existing urban parcels in 2012 using the framework of AICP (automatic identification and characterization of parcels) proposed by Long and Liu (2013). The vector CA module was applied for simulating urban expansion during 2012-2017. This module was examined using calibrated parameters abstracted from Beijing data. Three urban expansion scenarios - baseline, urban agglomeration, and new urban development- have been simulated during 2012-2017 by MVP-CA, respectively. The simulation results are shown in Figure 6. We validated the simulation results using two approaches, the first was to compare the baseline scenario of Beijing with the results using a raster CA model BUDEM we developed previously, and the second was performed by crowding validation.

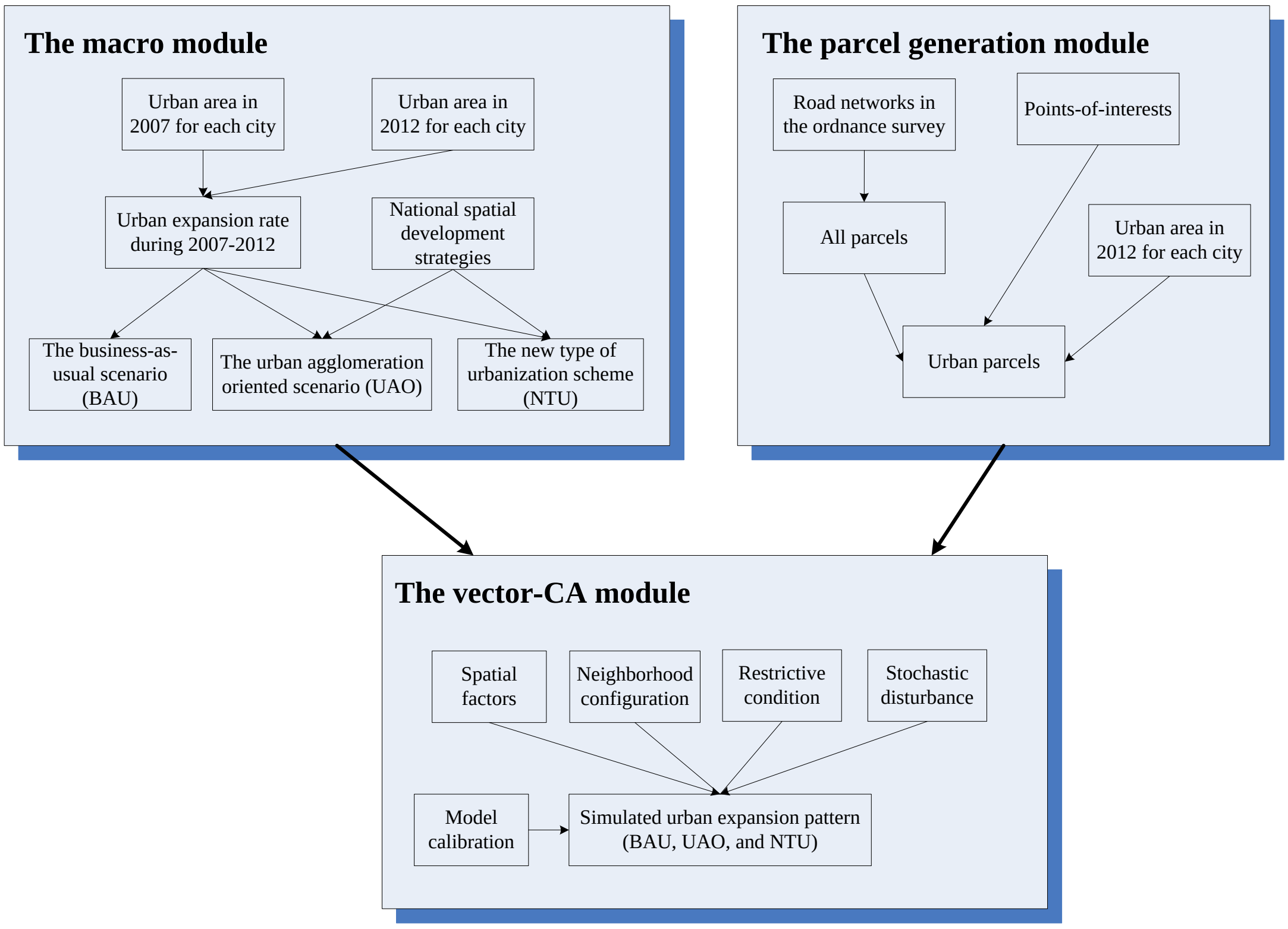

Figure 5 The structure and flow diagram of MVP-CA

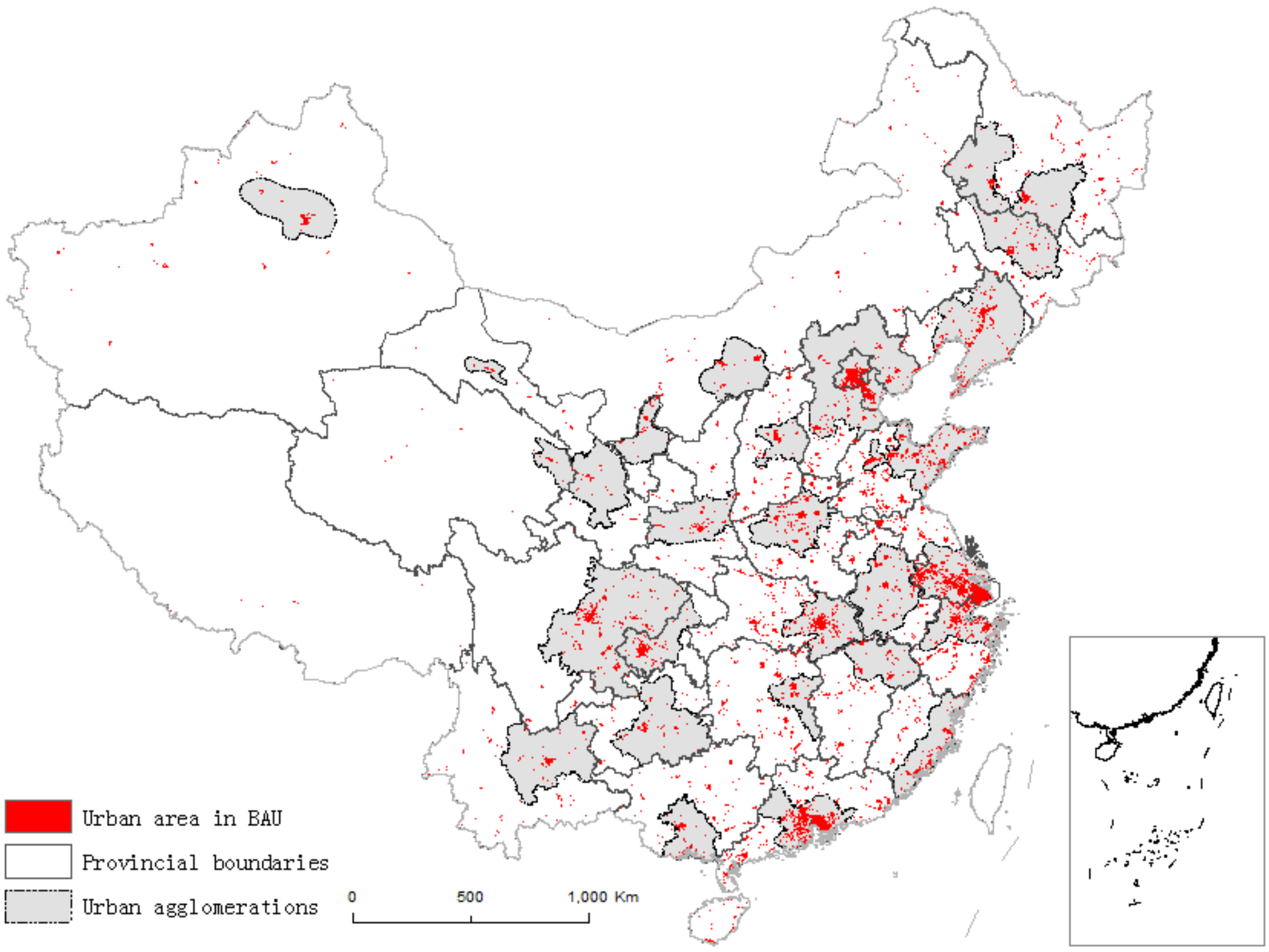

(a)

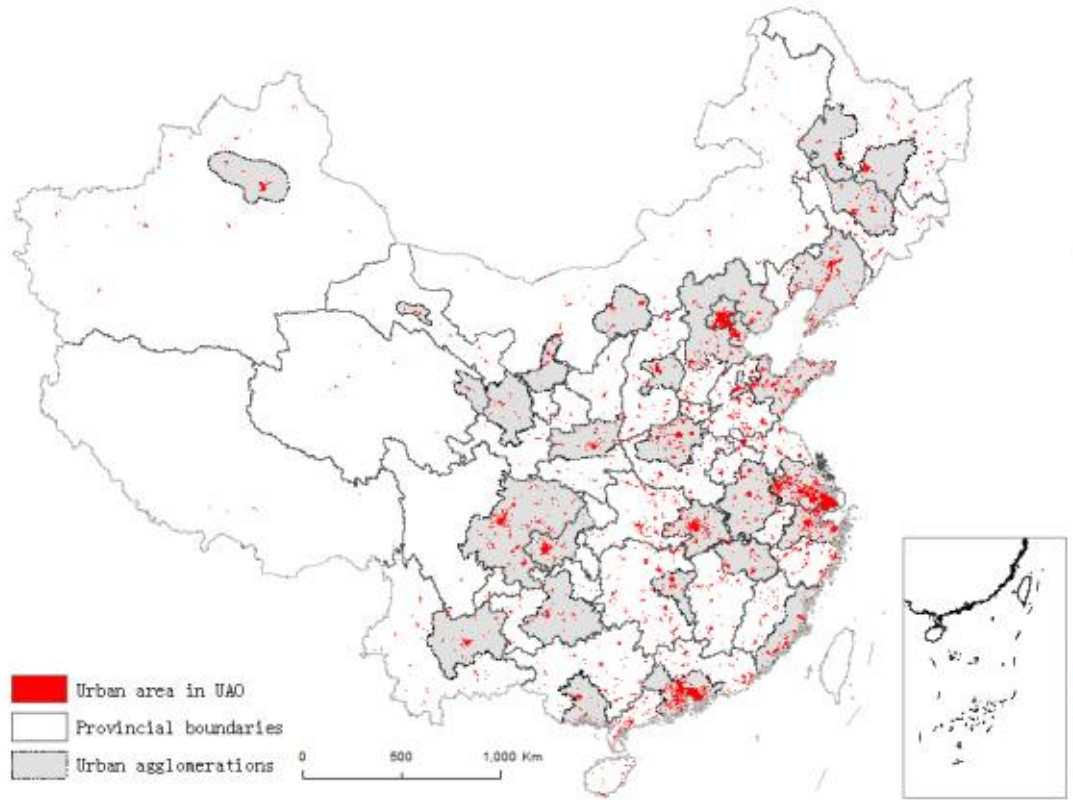

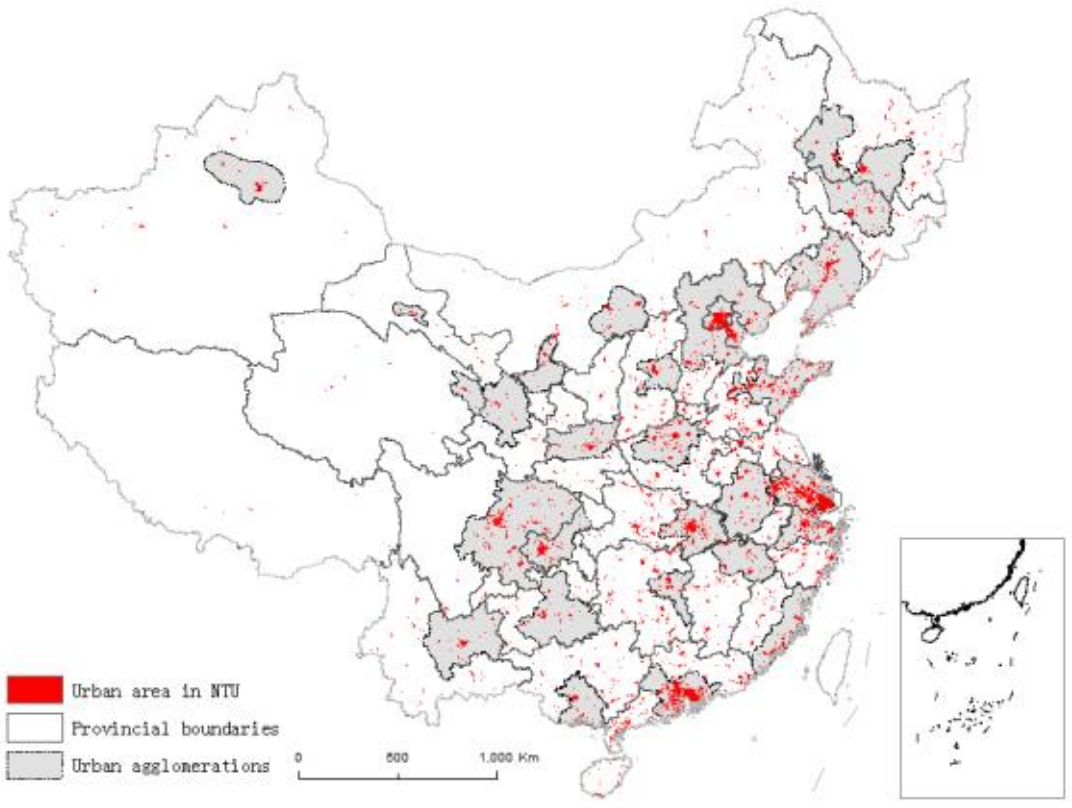

(b) (c)

Figure 6 Urban area of all Chinese cities (a), and urban expansion patterns of the entire China for three scenarios (a: BAU, b: UAO, c: NTU)

As the first large-scale urban expansion model in the fine-scale for the whole China, our contributions of this chapter mainly lie in the following aspects. First, a vector-based cellular automata model was introduced for simulating urban expansion in a super large geographical scale at the parcel level, which is rare in existing literature in the domain urban expansion modelling. Second, we proposed a solution for linking spatial development strategies with urban expansion via reflecting as the urban expansion speed of each city. This enables simulating macro policies in a very fine-scale through the channel of the MVP-CA model. Last, we simulated the near-future urban area for all Chinese cities in China, which, together with existing urban area, has already been shared online as an important data infrastructure for both practitioners and researchers.

## 3.3 Other ongoing projects by BCL

### *3.3.1 Estimating population exposure to PM2.5*

Chinese cities have for many years suffered from air pollution, which has been a major downside to rapid economic growth and increased urbanization. Currently, few studies of air pollution have been conducted to assess population exposure to PM2.5 over large geographical areas and time periods in China. The existing studies mainly focus on air pollution's effects on health and ecosystems or relevant monitoring methods and measurement, but less has been done on the link between urban spatial structure and air pollution exposure, not to mention their spatiotemporal pattern.

In this study, we collected daily PM 2.5 concentrations during April 08, 2013 and April 07, 2014 from 945 monitoring stations in 190 cities across China[2]. The air quality data were acquired from China National Environmental Monitoring Center (http://www.cnemc.cn). These datasets enable us to understand the PM 2.5 concentration of each station all year round, and can be used as a key input for our estimation. Considering the sparse distribution of monitoring stations across China, we further used MODIS AOD product to supplement the PM 2.5 estimates on a daily basis. Demographic statistics were drawn from China's 2010 census data. The spatial distribution of population density across China was determined by geocoding population density of each sub-district based on Google Map API. In total, there are 39,007 sub-districts[3] in China, and the average population density for all sub-districts is 977 persons per $km^2$. Population have been divided into three age groups (age 0-14, age 15-64, and 65 years and older), with an aim to differentiate the exposure estimates for different sensitive groups such as children and seniors. It is worth mentioning that this is the first time to use sub-district population density for estimating human exposure to air pollution in China, whereas former studies were conducted at the county level at best.

The population exposure estimation involves three major steps. **(1)** Interpolate the PM2.5 concentration site data into surface data using both ground station-level data and MODIS ADO: PM2.5 concentration data were obtained from all air quality monitoring stations across the country and supplemented with MODIS ADO data. Using numerous spatial interpolation methods, the station-level data can be interpolated into surface data. The outcome of this step is the average daily PM2.5 concentration over the entire area and over time. **(2)** Estimating population exposure to PM2.5 for each sub-district. Based on interpolated PM2.5 data, a daily PM2.5 concentration above the national standard of $75mg/m^3$ is considered to be unhealthy and thus defined as "exposed". In this way, the total exposed days all year round of each sub-district can be estimated. Further, the exposure intensity for each sub-district can be calculated using the

---

[2] There are 657 cities in mainland China as of the end of 2012.

[3] There are three forms of township-level administrative units in China, sub-districts (*jiedao*), towns (*zhen*), and township (*xiang*). Jiedaos are mainly in city area. *Jiedao*'s counterparts in the rural area are towns and townships. Hereafter in this chapter, we use the term sub-district for representing all types of township-level administrative units in China.

Equation: Exposure intensity = Population density * Exposed days. The greater exposed days or population density for a sub-district, the higher exposure intensity. This indicator reflects the strength of population exposure to PM2.5. The population density can be subject to specific sub-population groups for estimating the effects on members of sensitive groups. **(3)** Aggregating the estimated results spatiotemporally. To gain ideas on spatiotemporal pattern of population exposure to PM2.5, we can further aggregate the estimated results in both temporal and spatial dimensions. For the temporal dimension, the total number of exposed month can be calculated for each sub-district, thus presenting a big picture of population exposure to air quality over time. For the spatial dimension, the exposure of each city can be inferred by averaging the estimation results of all sub-districts in each city's administrative boundary.

The number of months subject to exposed condition across the entire country is presented in Figure 7.

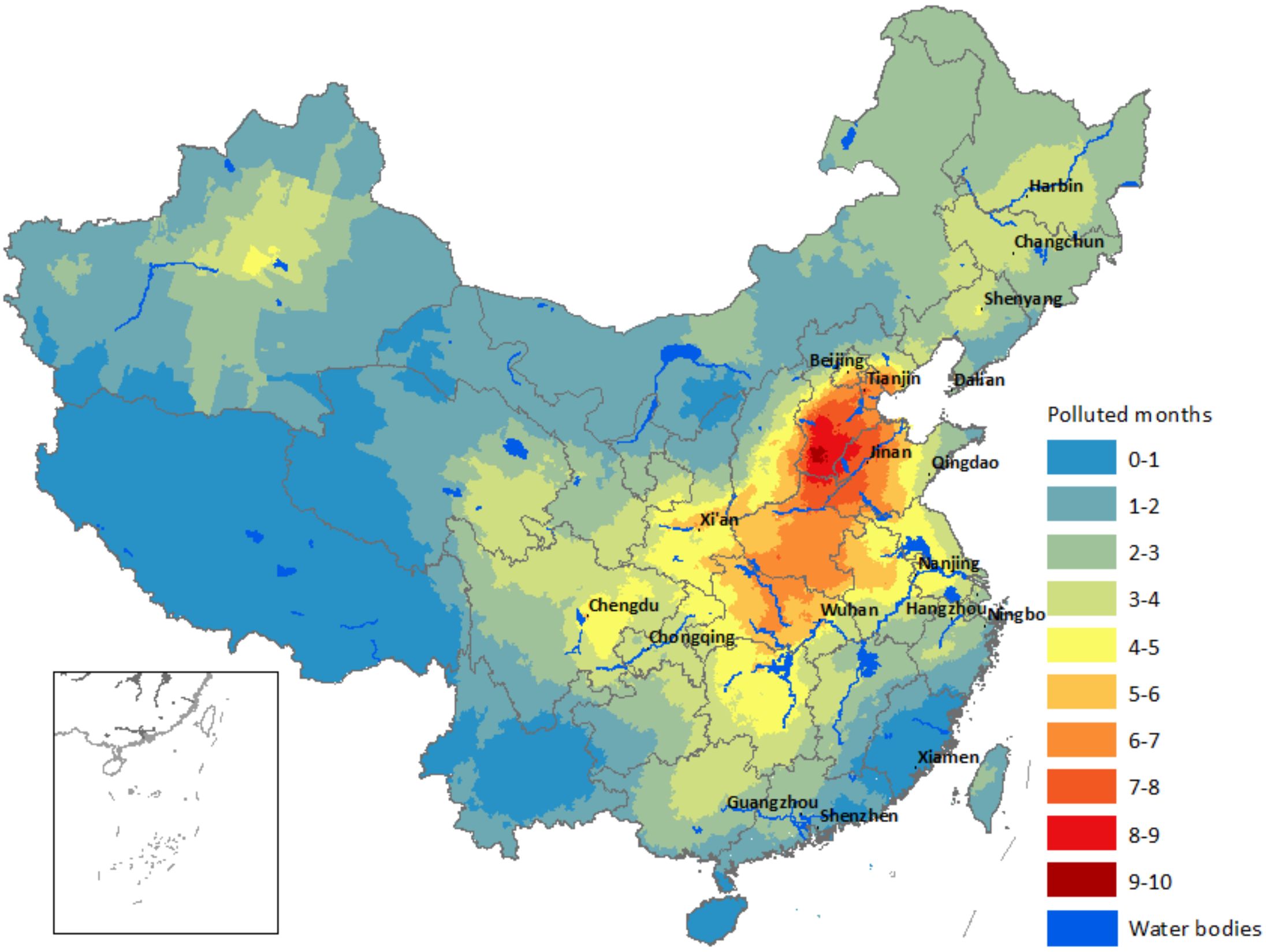

Figure 7 The number of total exposed months for each sub-district in China

The daily exposure for each sub-district was further aggregated by each month. Table 1 displays the percentage of exposure days per month from April 2013 to March 2014.

Table 1 Exposed days in each month for each sub-district in China

| Apr 2013 | May 2013 | Jun 2013 | Jul 2013 | Exposed |
|---|---|---|---|---|

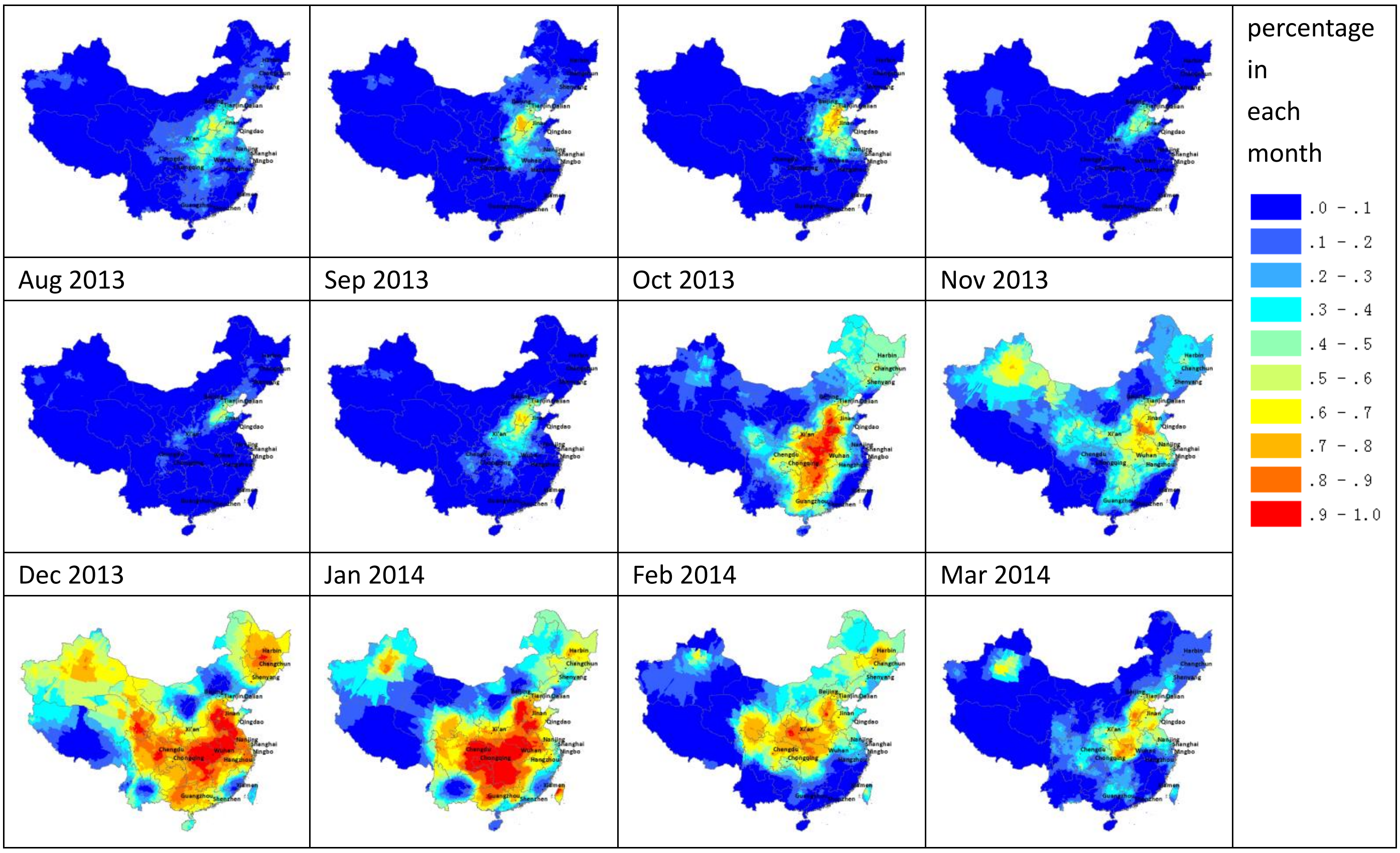

The exposure intensities were obtained by multiplying population density for each sub-district with the estimated exposure days during the period. The final result is presented in Figure 8. It is worth pointing out that the overall exposure intensity pattern generally coincides with the distribution of population density across the country.

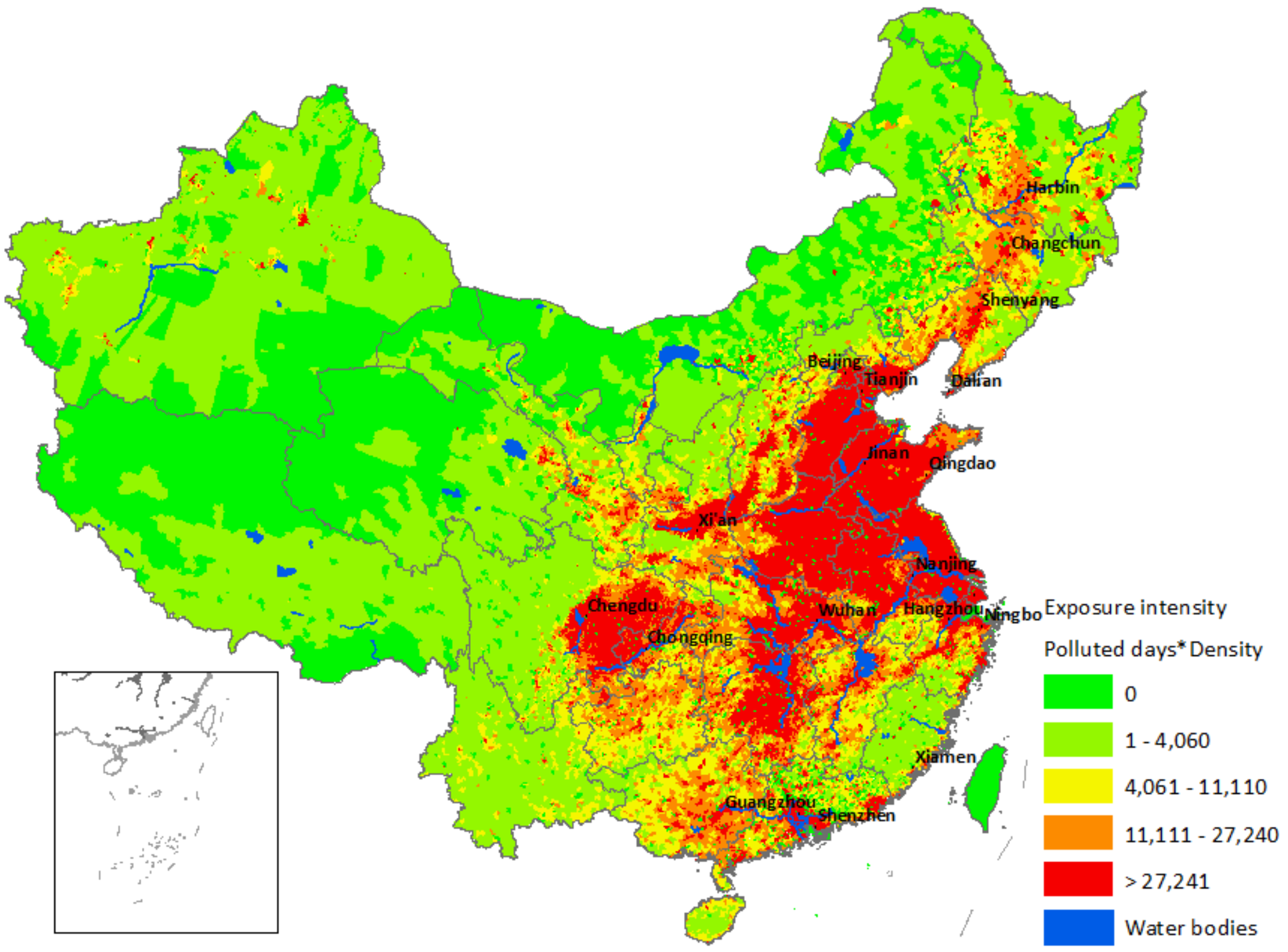

Figure 8 Exposure intensity in the town level of China (# days pop per km$^2$)

### *3.3.2 Evaluating urban growth boundaries for 300 Chinese cities*

Among the various urban growth management policies, urban containment policies have been widely adopted in an attempt to control the spread of urban areas, increase urban land use density, and protect open space (Nelson and Duncan, 1995, Long et al, 2011). In general, urban containment policies seek to manage urban growth through at least three different types of tools – greenbelts, urban growth boundaries (UGBs), and urban service boundaries (USBs) (Pendall et al., 2002). UGB is one of the most widely discussed tools in the planning field. Through zoning, land development permits, and other land-use regulation tools, UGBs demarcate urban and rural uses and aim to contain urban development within the predefined boundaries (Pendall et al., 2002). In China, urban construction boundaries determined in master or detailed plans have been commonly recognized as Chinese/planned UGBs (Long et al 2013), since they have a similar mechanism to UGBs in the U.S. as well as some other Western countries.

In China, conventional methods of delineating UGBs are based on planners' expertise and experiences; thus, they lack an adequate scientific basis and quantitative support. Consequently, the UGBs often fail to manage urban growth. According to Han et al. (2009)'s study on the examination of the implementation of planned UGBs within the sixth ring road of Beijing using multi-temporal remote sensing images, more urban land developments were found outside than inside the UGBs during the previous two planning periods (1983 to 1993 and 1993 to 2005). Tian and Shen (2011) and Xu et al. (2009) also suggested that substantial urban development occurred outside of UCBs in Guangzhou and Shanghai in recent years. These findings were also supported by Long et al (2012)'s research, which evaluated five master plans compiled and implemented in Beijing during 1958-2004. Though considerable progress has been made in revealing and quantifying the extent of urbanization and/or evaluating the urban policies' effectiveness on managing urban growth, we have found that most of them have been focused on a single city or region, and little work on the city-level comparison of the performance of UGB's implementation has been done.

Driven by our proposed urban growth simulation model and other relevant big models studies, we launched effort to create a systematic approach to horizontally examine and evaluate the effectiveness of UGBs across cities and regions. We collected raw planning drawing maps on planned UGBs in over 300 Chinese cities (see Figure 9 for a partial sample of cities) and digitalized the boundaries in GIS to facilitate spatial analysis and statistics on these planned UGBs. After that, the planned UGBs of a city were overlaid and compared with the actual extent of urban expansion in the past years since the plan was first implemented, and the ratio of legal development to all urban development can be directly calculated to facilitate city-level comparison. Furthermore, the ambitious degree of each city can be inferred by dividing the actual extent of urban expansion by the planned-to-be-development land area.
Compared with previous studies on big models to date, this research generalizes the planned UGBs across cities and regions and helps make sense of differing results of urban development. In addition, it can provide an insight of the overall trend of urban development in China and thus would be useful for planners to evaluate, monitor, and manage urban planning efforts.

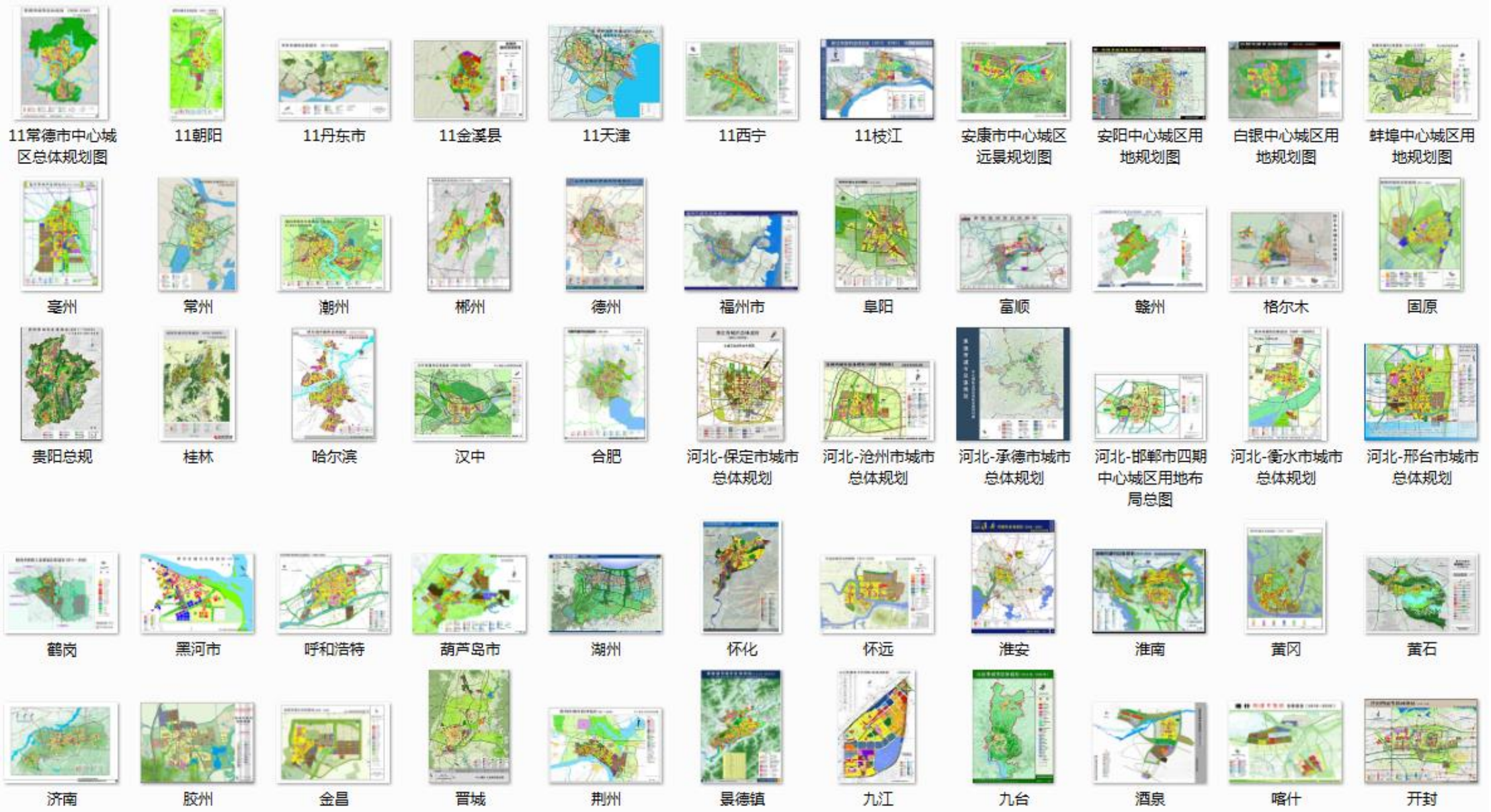

Figure 9 The profile of raw figures for planned UGBs (partially shown)

Meanwhile, the digitalized UGBs can also be used to supplement the MVP-CA urban expansion model for all Chinese cities (see our first case study in this chapter) as an institutional constraint, thus accounting for the simulation results. In addition, the project may help identify some universal law of governing the pattern of planned UGBs among all Chinese cities.

### *3.3.3 Population spatialization and synthesis for all Chinese cities*

Demographic and socio-economic information are important factors to incorporate in applied urban models. A number of studies have been conducted to generate synthetic individual data based on reweighting large-scale survey samples (Wu et al, 2008). In developing countries like China, demographic data at a fine-scale level is not universally available, nor are large-scale surveys for population synthesis. China is also facing an institution-induced digital divide –that is, a gap between data available to research and data gathered through official channels - as the government exercises tight control over the official data. This situation is common among developing countries (Tatem and Linard, 2011). We hereby aim to mitigate this gap by illustrating how the collection, analysis, and visualization of big (open) data can contribute to the progress of synthesizing micro data in developing countries.

In this study, we proposed an automatic process using open data for population spatialization and synthesis. Specifically, road network in OpenStreetMap was used to identify and delineate parcel geometries, while crowd-sourced POIs were gathered to infer urban parcels with a vector cellular automata model. Housing-related online check-in records were then applied for selecting residential parcels from all identified urban parcels. Finally, the released sub-district level population census, from which the distribution of and relationship among attributes are provided, was processed for synthesizing population attributes using a previous developed tool named "Agenter" (Long and Shen, 2013). Figure 10 presents the results of Beijing, which have been validated with ground truth manually-prepared dataset by planners in Beijing Institute of City Planning.

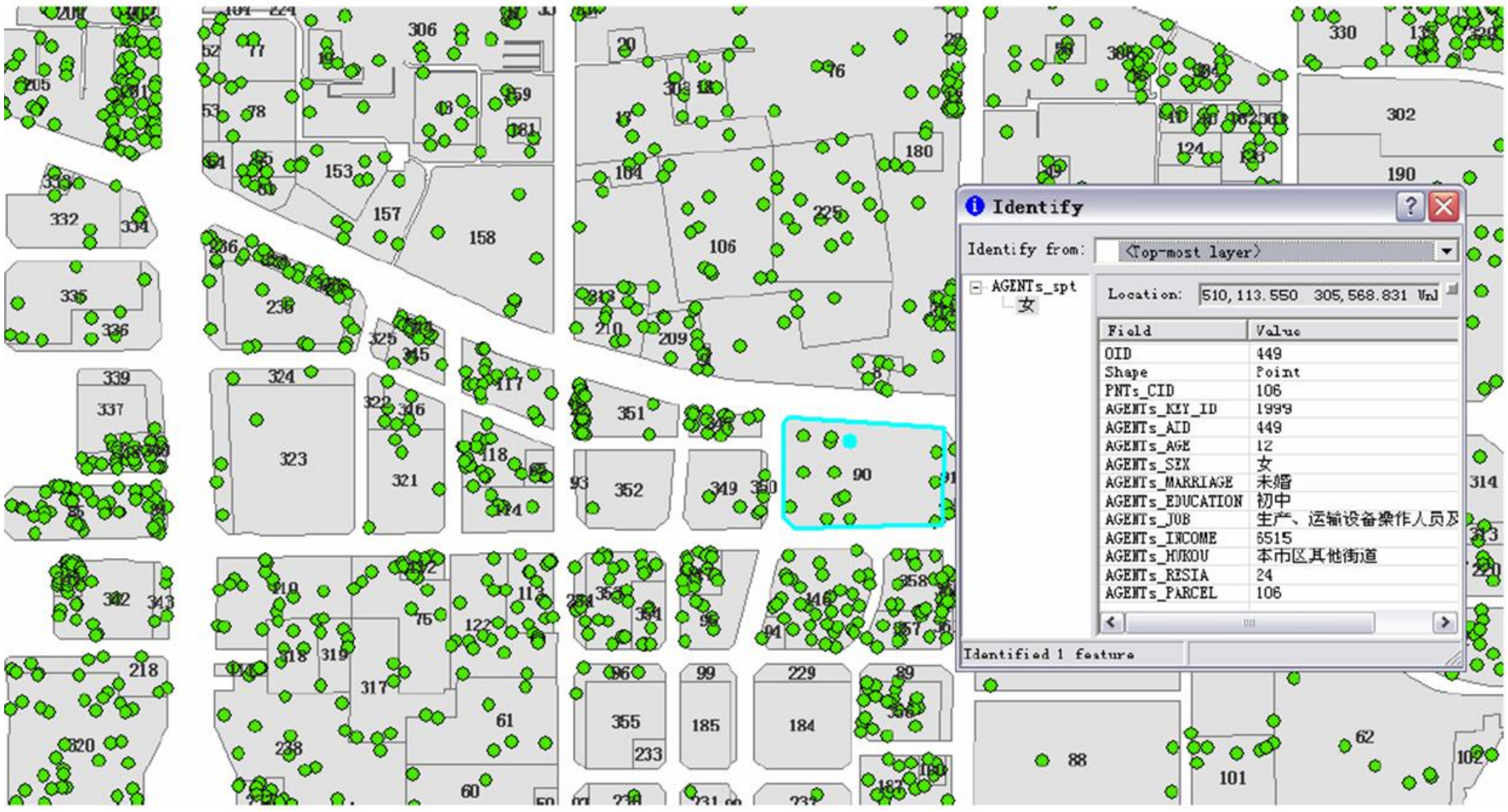

Figure 10 The spatial distributions of disaggregated agents in the Beijing Metropolitan Area (BMA) (partially shown)

The final product of our project is a dataset containing fine-scale resident distribution in space and their associated attributes throughout the country. This dataset can contribute to, but not limited to, the following areas. Firstly, the parcel-level population density dataset can serve as a powerful base for urban planners and researchers to analyse and solve a wide range of social and development related issues, such as quality-of-life, air pollution exposure, and population-based urban agglomeration. Secondly, the dataset we generated can act as a direct input in a form of spatial agents for the ever emerging agent-based models that previously had to utilize somewhat coarse dataset as inputs under the data-sparse conditions in China. Lastly, the generated results display its potential applications in market analysis, e.g. evaluating potential market for retails within the catchment area.

### *3.3.4 Identifying functional urban area for Chinese cities*

Since the introduction of the concept of metropolitan area (or functional urban area, FUA), the statistics agents and scholars from different countries have come up with different kinds of methods to delineate the boundary of metropolitan areas. One basic reason for identifying metropolitan areas is that, in reality, the administrative boundary of a city cannot well represent the actual size and impact of labor force and economic activity happening there. For instance, there are a large number of residents who are living in Yanjiao and Sanhe in Hebei Province but commuting to central Beijing for their job, and such commuting pattern would inevitably generate substantial influences on economy, housing, and environment that frequently take place beyond the traditional administrative boundary of Beijing. As a result, it becomes necessary to delineate boundaries of these geographic areas –that is, to define a metropolitan area – to capture its real population or economic functions. In general, a metropolitan area comprises a densely populated urban core and its less-populated neighboring territories that are socio-economically tied to the urban core. Though there has been no significant change in terms of the basic concept of metropolitan areas, their definition, criteria, and measurements are not consistent across countries and over time.

The National Bureau of Statistics of China has not published a standard for the delineation of metropolitan areas nationwide. However, some Chinese scholars have started to develop mathematical or econometric methods to define the metropolitan areas since the 1980s (Yu and Ning, 1983, Zhou, 1987) and there have been several studies and efforts that have been put into practice. Their methods generally followed a similar core-hinterland scheme as describe above. Among the definitions of the core area, there have been disputes but almost all of them were intended to use the statutory urban districts as the urban core. To determine the interacted exurban areas, due to lack of commuting data nationwide, researchers had to employ alternative measures for identifying linkages between core and outlying zones. These alternatives range from bus ridership, industrial economic indicators, to road network. To date, most of these studies were applied to a limited number of regions. There has not been any form of unified delineation methodology that can be applied to the entire county.

In our research, we incorporated various large-scale and high-resolution data, such as check-in records from commercial social networking websites and bus routes and stops from public transportation query services, for a purpose of identifying FUAs for over 300 large-sized cities in China. Our hypothesis is that, if an adjacent county has high interaction with the central county, there should be a considerable large percentage of residents commuting to the city for jobs. We first used the check-in record datasets to test our hypothesis. We assumed that a location where a targeted user has kept at least five check-in records within one year is to be this individual's residential place, and a location where he or she has ever posted job-related check-in records to be a working place. To this end, a potential commute trip can be established by linking those identified residential and working places. We found that most of commute trips take place between two distant cities – a pattern that is not quite reasonable in the real world. A possible explanation is that social network users are more likely to post a check-in record when they are visiting someplace new rather than a place they are familiar with. As a result, this proposed method became invalid.

We are now shifting our focus to bus routes and stops datasets. We have collected 622,375 bus stops in 203 mid- to large- sized Chinese cities. Since the routes and schedules of local bus services are commonly well-established and can reflect a real commuting pattern within the area, we started testing the viability of using these bus service datasets for delineating the FUAs. This is the first time that an attempt to map FUAs for a large number of cities in China has been made, and we hope the process of this research and final products can help make benchmark for delineating metropolitan areas nationwide.

### *3.3.5 Evaluating the quality of urban agglomerations in China at three levels*

Urban agglomerations (also referred to as city regions) have been emphasized in China's 11th Five-Year Plan and 12th Five-Year Plan for National Economic and Social Development. According to the plans, urban agglomerations will be developed as the main body of urbanization as well as the basic terrain unit in participating international competition and international division of labor (Wu et al, 2013). The central government of China has ratified and agreed to support more than 30 regional plans and development policies regarding urban agglomerations that involve 23 provinces, autonomous regions and municipalities. The objectives of these plans and policies are to guide the development of these urban agglomerations towards a healthier and more sustainable future.

The most accepted deliberation of urban agglomerations, which were proposed by Fang et al (2013), is presented in Figure 11. In total, there are 23 urban agglomerations, covering 355 cities. Of these identified urban agglomerations, none have been officially approved except Yangtze River Delta, Pearl River Delta, and Beijing-Tianjin-Hebei.

There is a lack of institutional knowledge and evidences to justify how these urban agglomerations have been developed and what are their rankings with respect to population, urban development, economic output, etc. In this study, we employed Big Models for evaluating these urban agglomerations (UAs) at three levels. First, we evaluated UAs morphologically using population density from the 2010 population census of China at the township level. Second, we evaluated UAs using district/county level urbanization ratio morphologically. Last, we evaluated UAs functionally using city-level transport flows extracted from inter-city trains, coaches, and flights. In addition to the examination of the degree of existing development within these UAs, the MVP-CA model developed in Section 3.2 have been applied for estimating the development potential for each UA in the next five years.

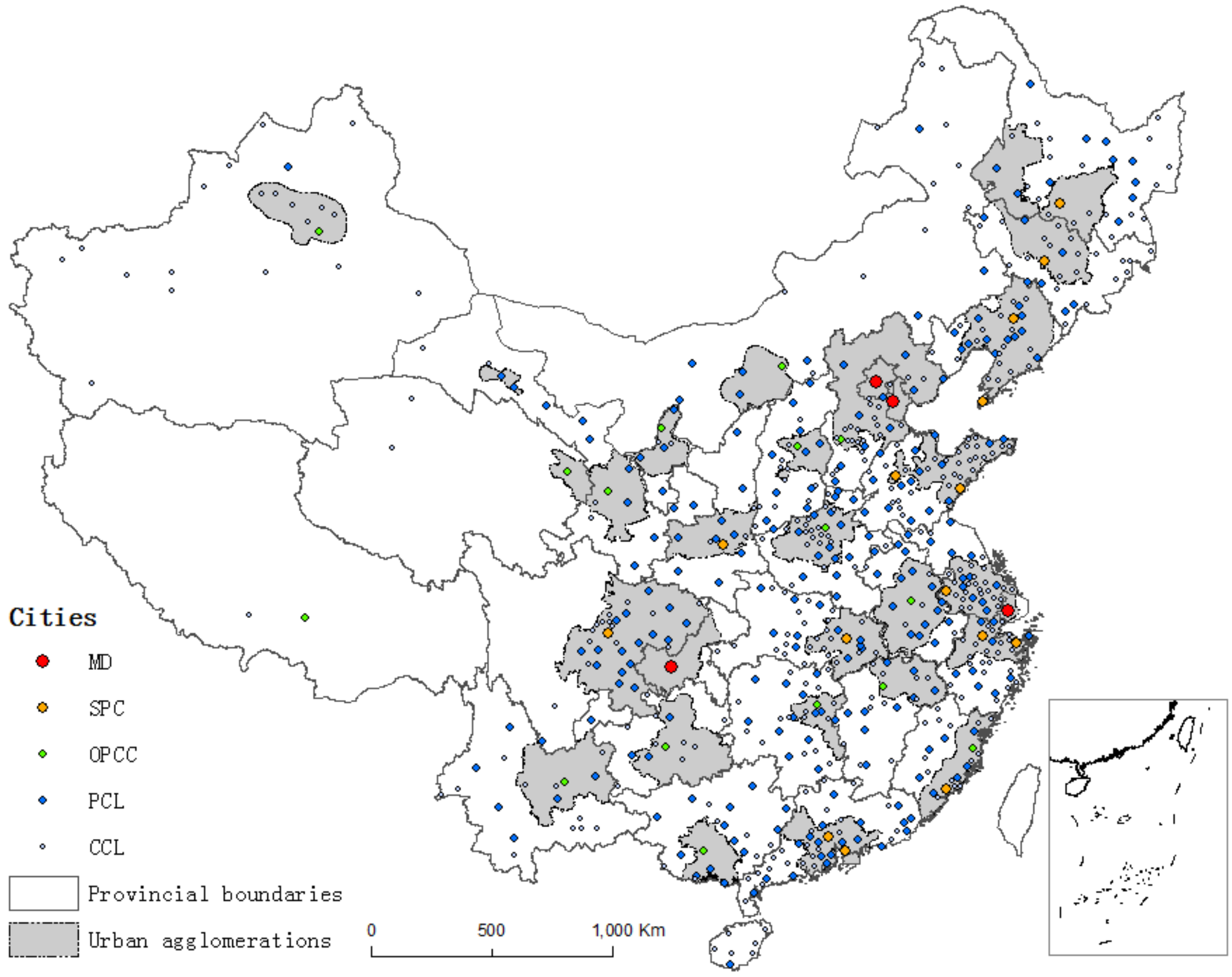

Figure 11 Urban agglomerations and cities in China

## 4 Conclusions and future directions

This chapter has proposed the concept of big model as a novel research paradigm for regional and urban studies. The concept, characteristics, and potential applications of big models have been elaborated. Meanwhile, we addressed several case studies to illustrate the progress of research and utilization on big models, including mapping urban areas for all Chinese cities, performing parcel-level urban simulation, and several ongoing research projects. Most of these applications can be adopted across the country, and all of them are focusing on a fine-scale level, such as a parcel, a block, or a township (sub-district), which is quite different from the existing studies using conventional models that are only suitable for a certain single or two cities or regions, or for a larger area but have to significantly sacrifice the data resolution. Believing that big models will mark a promising new era for the urban and regional study in the age of big data, we hope our efforts on urban analytics and modeling in Beijing City will set new research agenda and inspire innovative ideas all over the country.

There are several avenues on big models that deserve further studies. First, it is necessary to combine both intra-urban and inter-cities methods in big models. Existing case studies in this chapter mainly rely on bottom-up intra-urban approaches. City level linkages are essential to be included in big models in the next step. For instance, a spatial equilibrium module considering the provincial level input-output would replace the current “the macro module” in the near future. The integration of equilibrium mechanism with the dynamic CA model can link an inter-provincial - or even inter-city - simulation at the macro level with an urban expansion simulation at the local level. Second, more general theory on big models can be conceptualized through more in-depth case studies analysis.

**Acknowledgement**: We thank Miss Yichun Tu for her editing the language of this chapter.